\documentclass[twocolumn,prl]{revtex4}

\usepackage{graphicx}
\usepackage{dcolumn}
\usepackage{bm}

\newcommand{\be}{\begin{equation}}
\newcommand{\ee}{\end{equation}}
\newcommand{\ba}{\begin{eqnarray}}
\newcommand{\ea}{\end{eqnarray}}

\begin{document}
\title{Ultra-High Energy Cosmic Rays and the GeV-TeV Diffuse Gamma-Ray Flux}
\author{Oleg~E.~Kalashev$^a$, Dmitry~V.~Semikoz$^{a,b}$, G{\"u}nter Sigl$^b$}
\affiliation{$^a$ INR RAS, 60th October Anniversary pr. 7a, 117312 Moscow, Russia.\\
$^b$ APC, 10, rue Alice Domon et L\'eonie Duquet, Paris 75205, France.}

\begin{abstract}
Ultra-high energy cosmic ray protons accelerated in astrophysical
objects produce secondary electromagnetic cascades during propagation
in the cosmic microwave and infrared backgrounds. We show that
such cascades can contribute between $\simeq$1\% and $\simeq$50\%
of the GeV-TeV diffuse photon flux measured by the EGRET experiment.
The GLAST satellite should have a good chance to discover this flux.
\end{abstract}


\maketitle

{\it Introduction.}
Recently the HiRes collaboration established~\cite{hires_spectrum}
the existence of the GZK cutoff~\cite{gzk}.
If confirmed, this result suggests an astrophysical origin of
ultra-high energy cosmic rays (UHECR).

There are two important contributions to secondary electromagnetic (EM)
cascades from UHECR. One comes from the GZK process of pion production in 
interactions of UHECR protons with cosmic microwave (CMB) photons.
Most of the energy transferred to photons, electrons
and positrons in the subsequent pion decays would cascade down to
GeV-TeV energies, at which the Universe is transparent to photons.
If the spectrum of primary protons is a power law
$\propto E^{-\alpha}$, pion production energy losses and thus the
energy deposited into EM cascades increases with decreasing power
law index $\alpha$.

The second source of EM cascades is pair production
by protons on low energy photons, $p + \gamma\rightarrow p + e^+ +  e^-$.
This process is more efficient for steeper injection spectra with
larger $\alpha$. Both processes together imply a minimal
secondary photon flux at GeV-TeV energies.

The goal of the present work is to study the contribution of EM
cascades from UHECR proton interactions with
background photons to the diffuse
$\gamma-$ray background in astrophysical scenarios.
We study the parameter space of UHECR
models which fit the HiRes energy spectrum with the GZK cutoff.
We show that the UHECR contribution to
the $\gamma-$ray flux is in the range 1-50 \% of the diffuse
flux measured by EGRET~\cite{egret_new}. Relatively high values of
this flux should enable the GLAST satellite~\cite{glast}
to disentangle it from other contributions such as
from starforming galaxies~\cite{star_forming}, starbursts~\cite{starburst},
large scale structure formation shocks~\cite{structure},
AGNs~\cite{AGN_total}, blazars~\cite{Stecker:1996ma}, and
$\gamma-$ray bursts~\cite{grb}.

{\it Modeling the primary proton flux.}
We parametrize the proton injection spectrum as
$dN/dE\propto E^{-\alpha} \,\theta (E_{\max}-E)$,
where $E_{\max}$ is the maximal proton energy and $\alpha$ is the
power law index for which we consider the ranges
$2\times 10^{20} ~{\rm eV}~ \le E_{\max} \le 10^{21}$ eV and
$2 \le \alpha \le 2.7$, respectively.

We also consider  the case of sources with variable
density and/or luminosity. We assume the comoving source density
to scale as $n(z) = n_0 (1+z)^m  \theta (z_{\max}-z) \theta (z-z_{\min})$
where $m$  parameterizes the luminosity evolution. We consider the
range $-2 \le m \le 4$, which practically includes all astrophysical
scenarios. The parameters $z_{\min}$ and $z_{\max}$ are the redshifts of the
closest and most distant sources, respectively.
We choose $z_{\min} \le 0.01$ to avoid a GZK cutoff more pronounced
than observed~\cite{hires_spectrum}. We fixed $z_{\max}=3$ which
is large enough to take into account cosmologically distant sources. 
We note that both the luminosity evolution index $m$ and the maximum
redshift $z_{\max}$ have similar qualitative influence on both
proton spectrum and GeV-TeV cascade fluxes.

For the rectilinear propagation of protons and cascades of secondary electrons,
positrons and photons we used two independent codes~\cite{code1,code2},
which we compared on the level of individual interactions. For pion
production by protons and neutrons both codes use the SOPHIA
generator~\cite{sophia}. In addition, protons loose energy due to
production of $e^+e^-$ pairs, while neutrons decay. Secondary photons
produce single and double pairs on the low energy photon background.
Electrons and positrons interact via inverse Compton scattering and
triplet pair  production, and undergo synchrotron energy losses
in extragalactic magnetic fields.
Due to all those reactions secondary photons, electrons and positrons
cascade down to TeV energies, where they are not affected by interactions
with the CMB any more. However, even at these energies they can
still interact with infrared (IR) and optical photon backgrounds. 
We use the recent model of Ref.~\cite{IRO} for these backgrounds.

Highly structured sources and large scale magnetic
fields can lead to enhanced synchrotron fluxes up to TeV 
energies~\cite{Armengaud:2006dd}. To be conservative we neglect
magnetic fields and 3 dimensional effects here.

We fit the HiRes spectrum with the method described in Ref.~\cite{method}.
Among all models characterized by $m$, $\alpha$, $E_{\max}$, $z_{\min}$,
and $z_{\max}$ we choose those which fit the latest HiRes 
spectrum~\cite{hires_spectrum} at the 95 \% confidence level,
taking into account empty bins above the highest energy events observed,
as well as an energy uncertainty $\Delta E/E = 17$ \%, which
influences the shape of the spectrum around the GZK cutoff.  

We consider the two main scenarios for the transition from a cosmic
ray flux dominated by galactic sources to one dominated by extragalactic
sources: In the "dip scenario" extragalactic protons
dominate over the galactic contribution down to a few $10^{17}\,$eV
and the dip observed in the energy spectrum between
$\simeq1$ and $\simeq10\,$EeV is caused by pair production by
these protons~\cite{dip}.
We require predicted fluxes in this scenario fit the HiRes spectrum for
$E \ge 2\,$EeV.

In the second scenario, the "ankle" in the spectrum around
$\simeq5\times10^{18}\,$eV is due to a cross-over from low-energy
galactic to high-energy extragalactic cosmic rays.
Recent versions of this model include a mixed composition of UHECR 
at the highest energies~\cite{mixed_model}. To be conservative,
in this case we fit the HiRes spectrum only for $E \ge 40\,$EeV.  

\begin{figure}[h]
\includegraphics[height=0.5\textwidth,clip=true,angle=270]{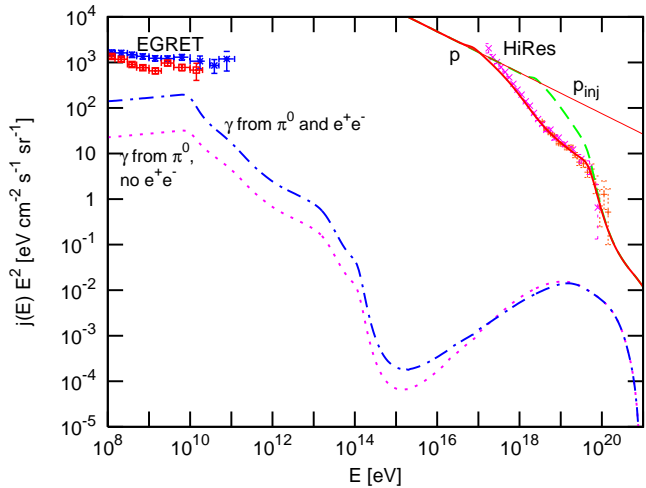}
\caption[...]{Primary proton and secondary $\gamma-$ray fluxes for
an injection spectrum $\propto E^{-2.45}$ up to $10^{21}\,$eV (shown as red,
solid straight line marked $p_{\rm inj}$) which
evolves as $(1+z)^3$ between $z_{\rm min}=0$ and $z_{\rm max}=3$.
In this case the UHECR flux is dominated by extragalactic 
protons down to a few $10^{17}\,$eV~\cite{dip}.
The lower red (solid) line is the proton flux, the blue
(dash-dotted) line is the corresponding secondary $\gamma-$ray
flux. The green (dashed) line is the proton flux without $e^\pm$ 
production and the magenta (dotted) line is the corresponding 
$\gamma-$ray flux. The UHECR flux observed by HiRes~\cite{hires_spectrum}
and two estimates of the extragalactic diffuse $\gamma-$ray background
deduced from EGRET data as blue (higher)~\cite{egret}
and red (lower) crosses~\cite{egret_new} are also shown.}
\label{fig1}
\end{figure}

{\it The Diffuse GeV-TeV $\gamma-$ray flux.}
We now discuss the possible range of contributions of 
UHECR to the diffuse $\gamma-$ray flux in the EGRET band.
In Fig.~\ref{fig1} we show a scenario where the dip
is due to pair production by extragalactic protons. The lower
red (solid) line was fitted to the HiRes spectrum~\cite{hires_spectrum}
at energies $E \ge 2\,$EeV and the corresponding EM cascade flux
is shown as blue (dash-dotted) line. Fig.~\ref{fig1} shows that
practically all EM energy ends up in the GeV-TeV region.
By also showing the proton spectrum (green, dashed line) and the corresponding
cascade flux (magenta, dotted line) when pair production by protons is
neglected, Fig.~\ref{fig1} demonstrates that the GeV-TeV $\gamma-$ray flux
is dominated by pair production losses of protons in this scenario.

The flux in energy carried by a differential spectrum $j(E)$ is given by
$\int E^2j(E)d\ln E$. One can then see from Fig.~\ref{fig1} that the UHECR
energy lost to pair production by protons (energy flux difference between
the red, solid and the green, dashed line) appears as the
energy flux difference between the blue, dash-dotted and the
magenta, dotted line, up to the pair flux which is not shown.
Furthermore, the UHECR energy lost to pion production by protons
(energy flux difference between the two red, solid
lines) appears as the energy flux in the magenta, dotted line,
up to the neutrino flux which is not shown.

\begin{figure}[ht]
\includegraphics[height=0.5\textwidth,clip=true,angle=270]
{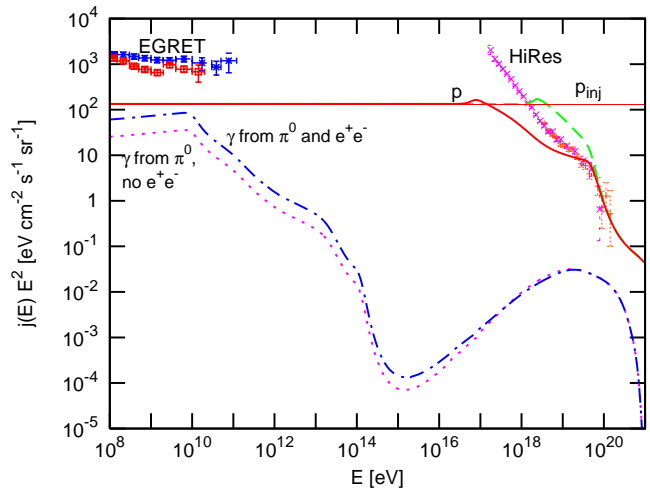}
\caption[...]{Same as  Fig.~\ref{fig1}, but for a scenario
with power law injection $\propto E^{-2}$ and evolution $\propto (1+z)^3$.}
\label{fig2}
\end{figure}

In Fig.~\ref{fig2} we show a scenario where the UHECR flux is
extragalactic only above the ankle. In the particular case $\alpha=2$
one can see the relation between the energy deposited in GeV-TeV photons
and the energy lost by protons more directly.
For this purpose we artificially continue the proton flux to low energies
where it has the same order of magnitude as the secondary photon flux.
Since protons loose similar amounts of energy to pion and to pair
production in this scenario, the contribution of these processes to the
GeV-TeV $\gamma-$ray flux is also comparable, contrary to the dip
scenario of Fig.~\ref{fig1}.

\begin{figure}[ht]
\includegraphics[height=0.5\textwidth,clip=true,angle=270]{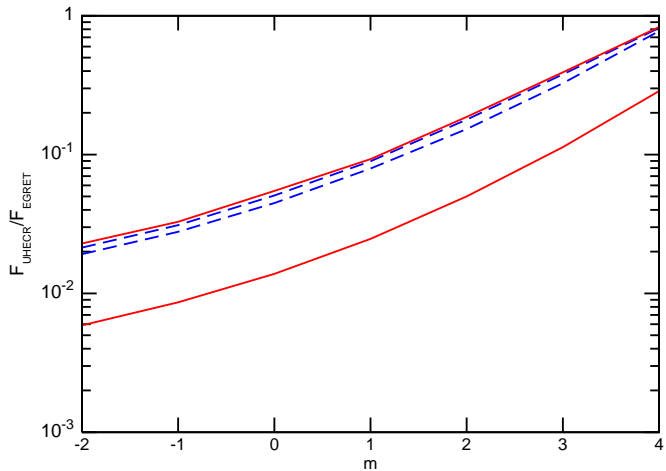}
\caption[...]{Dependence on redshift evolution index $m$ of the minimal
and maximal fractional
contribution of UHECR interactions to the EGRET flux between 1 and 2 GeV.
Blue, dashed lines are for fitting above 2 EeV
(dip scenario of the type shown in Fig.~\ref{fig1}) and
red, solid lines are for fitting the UHECR spectrum above 40 EeV.}
\label{fig3}
\end{figure}

In Fig.~\ref{fig3} we show the range of possible contributions of UHECR
interactions to the EGRET flux, as a function of redshift evolution index $m$.
We express this as the fraction of integral fluxes between 1 and 2 GeV,
where the EGRET energy flux $E^2 j(E)$ is minimal. Fig.~\ref{fig3} shows
that the diffuse $\gamma-$ray flux
in the EGRET band strongly depends on the source luminosity evolution
index $m$. In the ankle scenario, for a given value of $m$, it still depends
on other parameters within a factor 3, whereas in the dip
scenario the scatter is much smaller due to partial degeneracy between
$m$ and $\alpha$. All realistic astrophysical
source distributions have $m \ge 0$, which implies that the
contribution of secondary photons from UHECR will be at least $\simeq$1\%.
For stronger evolution, $m=3-4$, this fraction increases to more
than 50 \% of the EGRET flux.

\begin{figure}[ht]
\includegraphics[height=0.5\textwidth,clip=true,angle=270]{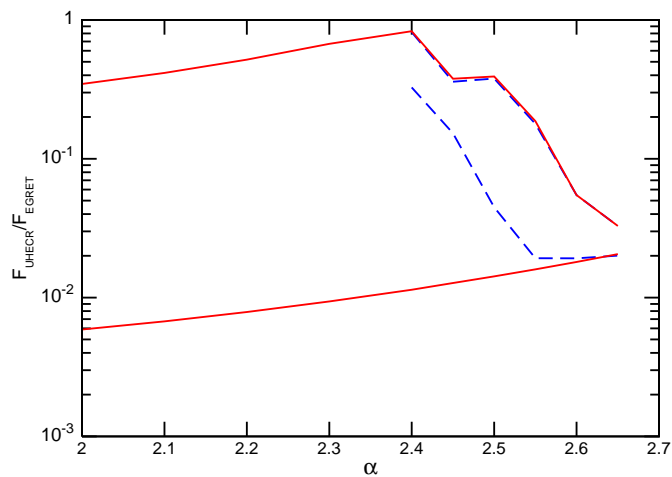}
\caption[...]{As Fig.~\ref{fig3}, but as function of the UHECR injection
spectral index $\alpha$.}
\label{fig4}
\end{figure}

In Fig.~\ref{fig4} we show the range of possible
contributions of UHECR interactions to the EGRET flux, as a
function of the UHECR injection power law index $\alpha$. Contrary to the
case of Fig.~\ref{fig3}, the scatter
is larger, especially for small values of $\alpha=2 - 2.4$. This is due
to the strong dependence of the flux in the EGRET band on the value
of $m$ for any given $\alpha$, see Fig.~\ref{fig3}.
The lower lines correspond to minimal values of $m=-2$, while the
maximum is defined by $m=4$ for $\alpha\lesssim 2.4$, and by smaller
values of $m$ for $\alpha \gtrsim 2.4$. Other parameter combinations
would overproduce the cosmic ray flux below $\simeq10\,$EeV.

\begin{figure}[ht]
\includegraphics[height=0.35\textwidth,clip=true,angle=0]{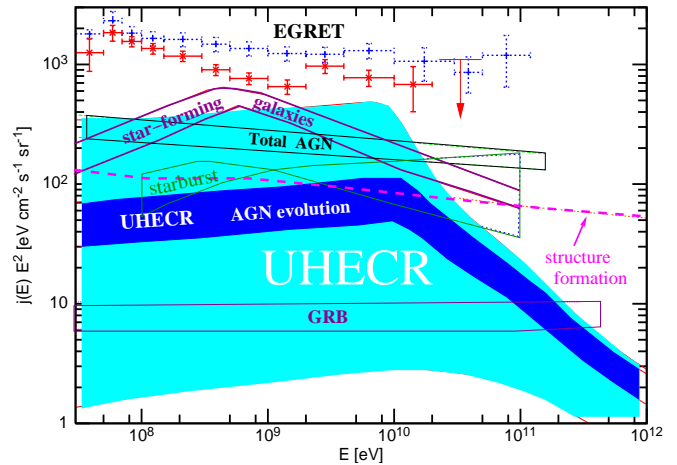}
\caption[...]{The possible range of UHECR induced cascade fluxes
(light shaded band) compared to estimated $\gamma-$ray fluxes directly
produced by starforming galaxies~\cite{star_forming}, starbursts~\cite{starburst},
large scale structure formation shocks~\cite{structure}, AGNs~\cite{AGN_total},
and $\gamma-$ray bursts~\cite{grb}. The dark shaded band shows the range
of EM cascade fluxes from UHECR sources evolving as AGN~\cite{Waxman:1998yy}.
The extragalactic diffuse
$\gamma-$ray background from EGRET is as in Fig.~\ref{fig1}.}
\label{fig5}
\end{figure}

In Fig.~\ref{fig5} we compare the range of EM cascade fluxes
from UHECR with other possible astrophysical contributions in
the EGRET band. Note that most of the uncertainty of the UHECR
cascade flux comes from the unknown source evolution. The scatter
for a given evolution such as for AGN is thus much smaller, as
seen from Fig.~\ref{fig5}. We remark that dark matter annihilations
may also contribute to the diffuse flux~\cite{Elsaesser:2004ap}.

The GeV-TeV cascade flux in scenarios where extragalactic cosmic rays 
dominate down to below the ankle at $\simeq5\times10^{18}\,$eV
would be practically diffuse for an instrument
such as GLAST which will be sensitive to anisotropies down to
$\sim0.1$\%~\cite{Hooper:2007be}. This is because the cosmic ray flux
below the ankle is dominated by cosmological sources and is isotropic
at the percent level, and because large scale magnetic fields should
lead to significant additional isotropization of primary 
protons~\cite{Sigl:2004gi} and secondary pairs.

In contrast, lacking statistics, UHECR anisotropies at
the 10\% level can currently not be ruled out at energies around or above
the GZK cutoff, $E> 40$ EeV, because sources of such UHECR may have a small density $\sim10^{-5}\,{\rm Mpc}^{-3}$~\cite{Kachelriess:2004pc}.
Furthermore, at energies $E\sim200\,$GeV--$10\,$TeV the $\gamma-$ray absorption
length in the IR background becomes small compared to the Hubble radius,
but still large compared to the $\sim10\,$Mpc length scales of pion
production and EM cascade development around the source. This is also
reflected in the amount of $\gamma-$ray flux suppression in
Figs.~\ref{fig1} and~\ref{fig2}. Independent
of the poorly known size of deflection, this could lead to
considerable correlation with nearby UHECR sources, as in the
case where discrete sources emit very high energy $\gamma-$rays
directly~\cite{Cuoco:2006tr}.
In scenarios such as in Fig.~\ref{fig2}, where the secondary cascade flux
is dominated by pion production due to relatively hard UHECR injection
spectra, this flux could, therefore, exhibit detectable
small scale anisotropy around $\sim200\,$GeV.

{\it Conclusions.}
UHECR interactions with low energy photons can significantly contribute to
the observed diffuse flux of
$\gamma-$rays at energies between $\sim$100 MeV and $\sim\,$TeV.
In this paper we studied the dependence of this contribution
on unknown parameters of astrophysical UHECR scenarios. We found that UHECR contribute
no less than 1\% to the observed EGRET flux, and up to 50\% in some cases.
This suggests that the GLAST satellite, which at GeV energies
will be $\simeq30$ times more sensitive to point sources than the EGRET
experiment, will likely be sensitive to the UHECR induced contribution.
Even ground-based instruments such as HESS and the future CTA 
may be sensitive to the cascade flux between $\simeq0.1$ and $\simeq10\,$TeV,
although such experiments are less well suited for diffuse backgrounds.

If sources of extragalactic highest energy cosmic rays are rare
and dominate the flux down to only $\simeq5\times10^{18}\,$eV, the
cascade background may have significant anisotropy at energies
around $200\,$GeV.

To summarize, future measurements of resolved and unresolved
components of the diffuse EGRET $\gamma-$ray background or upper limits
on such components can give important information on UHECR
origin and the distribution of their sources.

{\it Acknowledgments.}
We thank F.~Stecker for providing us with tables for the IR/optical
backgrounds for the model of Ref.~\cite{IRO}.
The numerical simulations were performed at the computer cluster of
the Theory Division of INR RAS. O.K. acknowledges financial support
from in2p3/CNRS for a collaboration visit at APC, Paris.



\begin{thebibliography}{99}

\bibitem{hires_spectrum}
  R.~Abbasi {\it et al.}  [HiRes Collaboration],
  arXiv:astro-ph/0703099.

\bibitem{gzk} K.~Greisen,
Phys.\ Rev.\ Lett.\  {\bf 16}, 748 (1966);
G.~T.~Zatsepin and V.~A.~Kuzmin,
JETP Lett.\  {\bf 4}, 78 (1966) [Pisma Zh.\ Eksp.\ Teor.\ Fiz.\  {\bf
4}, 114 (1966)].

\bibitem{egret}
P.~Sreekumar {\it et al.},
Astrophys.~J. {\bf 494},  523 (1998)
[astro-ph/9709257].

\bibitem{egret_new}
 A.~W.~Strong, I.~V.~Moskalenko and O.~Reimer,
  Astrophys.\ J.\  {\bf 613}, 956 (2004)
  [arXiv:astro-ph/0405441].

\bibitem{glast} For general information
see {\sf http://www-glast.stanford.edu}

\bibitem{star_forming}
   V.~Pavlidou and B.~D.~Fields,
   Astrophys.\ J.\  {\bf 575}, L5 (2002)
   [arXiv:astro-ph/0207253].

\bibitem{starburst}
   T.~A.~Thompson, E.~Quataert and E.~Waxman,
   Astrophys.\ J.\  {\bf 654}, 219 (2006)
   [arXiv:astro-ph/0606665].

\bibitem{structure}
   U.~Keshet, E.~Waxman, A.~Loeb, V.~Springel and L.~Hernquist,
   Astrophys.\ J.\  {\bf 585}, 128 (2003)
   [arXiv:astro-ph/0202318].

\bibitem{AGN_total}
   C.~D.~Dermer,
   arXiv:astro-ph/0605402.

\bibitem{Stecker:1996ma}
  F.~W.~Stecker and M.~H.~Salamon,
  Astrophys.\ J.\  {\bf 464}, 600 (1996)
  [arXiv:astro-ph/9601120].

\bibitem{grb}
   C.~D.~Dermer,
   arXiv:astro-ph/0610195.

\bibitem{code1}
  E.~Armengaud, G.~Sigl, T.~Beau and F.~Miniati,
  arXiv:astro-ph/0603675;
  see {\sf http://apcauger.in2p3.fr//CRPropa}.
  
\bibitem{code2}
 O.~E.~Kalashev, V.~A.~Kuzmin and D.~V.~Semikoz,
 arXiv:astro-ph/9911035;
  Mod.\ Phys.\ Lett.\  A {\bf 16}, 2505 (2001).

\bibitem{sophia}
  A.~Mucke, R.~Engel, J.~P.~Rachen, R.~J.~Protheroe and T.~Stanev,
  Comput.\ Phys.\ Commun.\  {\bf 124}, 290 (2000)
  [arXiv:astro-ph/9903478].

\bibitem{IRO} 
F.~W.~Stecker, M.~A.~Malkan and S.~T.~Scully,
  Astrophys.\ J.\  {\bf 648}, 774 (2006)
  [arXiv:astro-ph/0510449].

\bibitem{Armengaud:2006dd}
  E.~Armengaud, G.~Sigl and F.~Miniati,
  Phys.\ Rev.\  D {\bf 73}, 083008 (2006).

\bibitem{method}
  G.~Gelmini, O.~Kalashev and D.~V.~Semikoz,
  arXiv:astro-ph/0702464.

\bibitem{dip} V.~Berezinsky, A.~Z.~Gazizov and S.~I.~Grigorieva, 
 Phys.\ Rev.\ D {\bf 74}, 043005 (2006) [arXiv:hep-ph/0204357]; 
astro-ph/0210095;
Nucl.\ Phys.\ Proc.\ Suppl.\  {\bf 136}, 147 (2004) [astro-ph/0410650];
Phys.\ Lett.\ B {\bf 612} (2005) 147 [astro-ph/0502550].

\bibitem{mixed_model} 
D.~Allard, E.~Parizot and A.~V.~Olinto, 
Astropart.\ Phys.\  {\bf 27}, 61 (2007) [arXiv:astro-ph/0512345].  

\bibitem{Hooper:2007be}
  D.~Hooper and P.~D.~Serpico,
  arXiv:astro-ph/0702328.

\bibitem{Waxman:1998yy}
   E.~Waxman and J.~N.~Bahcall,
   Phys.\ Rev.\  D {\bf 59}, 023002 (1999)
   [arXiv:hep-ph/9807282].

\bibitem{Elsaesser:2004ap}
  D.~Elsaesser and K.~Mannheim,
  Phys.\ Rev.\ Lett.\  {\bf 94}, 171302 (2005)
  [arXiv:astro-ph/0405235].

\bibitem{Sigl:2004gi}
  G.~Sigl, F.~Miniati and T.~Ensslin,
  Nucl.\ Phys.\ Proc.\ Suppl.\  {\bf 136}, 224 (2004)
  [arXiv:astro-ph/0409098].

\bibitem{Kachelriess:2004pc}
  M.~Kachelriess and D.~Semikoz,
  Astropart.\ Phys.\  {\bf 23}, 486 (2005)
  [arXiv:astro-ph/0405258].

\bibitem{Cuoco:2006tr}
  A.~Cuoco, S.~Hannestad, T.~Haugbolle, G.~Miele, P.~D.~Serpico and H.~Tu,
  arXiv:astro-ph/0612559.

\end{thebibliography}
\end{document}